\documentclass[ reprint, amsmath, amssymb, aps, prd]{revtex4-1}

\usepackage[english]{babel}
\usepackage{amsbsy}
\usepackage{amstext}
\usepackage{amsmath}  
\usepackage{amssymb}
\usepackage{graphicx}

\usepackage{pst-node} 
\usepackage{diagbox}
\usepackage{multirow}
\usepackage{todonotes}

\makeatletter
\def\captionof#1#2{{\def\@captype{#1}#2}}
\makeatother
\begin{document}
\title[Constructive summation of the leading Quasi Normal from a population of BH]{Constructive summation of the (2,2) quasi normal mode from a population of black holes}
\author{ C. F. Da Silva Costa$^{1}$, S. Tiwari$^{2, 3}$,  S. Klimenko$^{1}$, F. Salemi$^{4}$}
\address{(1) University of Florida, Gainesville, Florida, USA }\email{filipe.dasilva@ufl.edu}
\address{(2) INFN, Trento Institute for Fundamental Physics and Applications, Trento, Italy}
\address{(3) Gran Sasso Science Institute (INFN), Via F. Crispi 7, I-67100, L'Aquila, Italy}
\address{(4) Albert-Einstein-Institut, Max-Planck-Institut f\"ur Gravitationsphysik, Hannover, Germany}
\date{\today}

\begin{abstract}{
The quasi normal modes (QNMs) associated with gravitational-wave signals from binary black hole (BBH) mergers can provide deep insight into the remnant's properties. Once design sensitivity is achieved, present ground-based gravitational wave interferometers could detect potentially hundreds of BBH signals in the coming years. For most, the ringdown phase will have a very weak signal-to-noise ratio (SNR). 
Signal summation techniques allow information extraction from the weak SNR ringdowns.

We propose a method to constructively sum the (2,2) QNM from different BBH signals by synchronizing and rescaling them. The parameter space adopted to test the method is presently limited to mass ratio $q\leq3$, initially non-spinning black holes with face-on orientation.
Moreover, since the synchronisation procedure fails for the weakest signals, we select all ringdowns with SNR above 2.6.
Under these conditions, we show that for different  BBH populations, 40 to 70\% of all the potential detections could be used for the summation while still ensuring a summed SNR of $\sim$80\% of the maximal achievable SNR (i.e. for ideally synchronized signals).}
\end{abstract} 
\pacs{04.80.Nn, 95.55.Ym}
\maketitle

\section{Introduction}
A binary black hole (BBH) is expected to form a perturbed Kerr black hole (BH) \cite{bib:kerr63}.  Its perturbations are damped oscillations~\cite{bib:Vishveshwara70}, which are the superposition of quasi normal modes (QNMs) \cite{bib:Press71, bib:QNM_sound}.  The no-hair theorem \cite{RuffiniWheeler71} tells us that a Kerr black hole can be described by two parameters, its mass, $M_{BH}$, and its dimensionless spin, ${a}$. These two BH parameters can be obtained from the QNMs and hence used to carry out tests of general relativity; for instance, test the Kerr nature of the black hole or a consistency test of general relativity (comparing QNM inferred parameter values with those derived from the inspiral-merger) \cite{Gossan1, bib:theGRtest}.

Presently, the rate of stellar mass BBH mergers is estimated to be 12-213 $Gpc^{-3} yr^{-1}$ \cite{bib:GW170104}; implying the possible detection of hundreds of BBHs in the coming years by GW interferometers \cite{bib:predictionLigo}. Most of these BBH signals are expected to have a weak ringdown where no information can be extracted. 
Indeed, considering the four LIGO observed BBH merger events: GW150914, GW151226, GW170104 and GW170814 \cite{GW150914, GW151226, bib:GW170104, bib:GW170814_limitedAuthors}, only GW150914 has a ringdown with high enough signal-to-noise ratio (SNR $\sim 7$) to extract information \cite{bib:theGRtest, TestingGR, PhysicsGW150914, PropertiesGW150914}.

For this reason, methods being developed to detect QNMs from BHs  \cite{ClaudilQNM, CaudillReducedBaisis, MeidamTigger, Berti2006, bib:dain} are targeting the more sensitive future  generations of ground and space-based detectors.
High SNR ringdown signals will be most likely rare, allowing informative general relativity consistency tests in only a few cases.  However, 
signal summation techniques  \cite{MeidamTigger, bib:BHspectroscopy} applied to most weak ringdown signals can help to extract information otherwise lost.

We have developed a method to constructively sum up the dominant (2,2) QNM from several BBH signals. The resultant signal is a ``normalized" (2,2) mode which could be used to infer the properties for the population of remnant BH's, i.e. ``normalized" mass, $M_{BH}^{\prime}$, and spin, $a^{\prime}$. This, in turn, can provide a weak test on the Kerr nature of the BH population.

The subdominant modes $lm$=(3,3), (2,1) and (4,4) will provide tighter constrains on the BH population's Kerr nature,
however, presently our methodology can not be applied to the subdominant modes. Such an extension of our methods would require the relative phases between these modes to synchronise them, which is currently unknown.

In Section \ref{sec:MethodDescription}, we detail the method and its limitations. Section \ref{sec:application} presents the results of the method tested with a population of simulated ringdowns. Finally, we present our conclusions in Section \ref{sec:Conclusion}. 

\section{Method description}
\label{sec:MethodDescription}
QNMs have a rich and complicated structure, as such, we choose to primarily test the present method with a reduced set of parameters. The present detections have a mass ratio, $q$, between 1 and 2, we therefore constrain $q\leq$3. We also limit our study to initially spinless BBHs with face-on orientation. Given these constraints, we choose the four SXS waveforms \cite{SXS:catalog, Mroue:2012kv, Mroue:2013xna, Buchman:2012dw, Lovelace:2010ne, Lovelace:2011nu} shown in Fig.~\ref{fig:ModesOfq3} and detailed in Appendix \ref{sec:waveforms}. 
 
The QNMs are described by:
 \begin{equation}
 h_{lm} = A_{lm} e^{-\pi f_{lm}/Q_{lm} t} \cos(2\pi f_{lm} t +\Phi_{lm})\, ,
\label{eqt:hwave_simple}
\end{equation}
where  $l,m > 0$ are the spheroidal harmonic indices with no overtones due to low amplitudes  \cite{bib:Ioannis1}; while $A_{lm}$ and $\Phi_{lm}$ are respectively the amplitude and phase of each mode.
Frequencies, $f_{lm}$, and quality factors, $Q_{lm}$, are related to the remnant black hole mass, $M_{BH}$, and the dimensionless spin, $a$, \cite{bib:Leaver85, Berti2006}; this is explained in more detail in Appendix \ref{sec:QNM}.  For the (2,2) mode, this reduces to,
 \begin{eqnarray}
f_{22} = \frac{1}{2\pi M_{BH}}\left[1.525 - 1.157 (1-a) ^{0.129}\right] \label{eqt:RingDown22f}\,,\\ 
Q_{22} = 0.700+1.419 (1-a)^{-0.499} \label{eqt:RingDown22q}\,, 
\end{eqnarray}
where we have set the speed of light in a vacuum and the gravitational constant, $c=G=1$.\\

In order to sum constructively all the (2,2) QNMs, the ringdown signals are rescaled so that they have the same $f_{22}$ frequency.
Their frequencies depend on the remnant BH mass and spin, which can be estimated by {\textit{LALinference}} using the inspiral-merger part of the signal  \cite{bib:robustParamters}.
Rescaled ringdown signals are then synchronized to the same time of reference and summed. 
The resulting signal is fitted to the (2,2) function given by Eq.~\ref{eqt:hwave_simple}, where the frequency and damping factors are substituted by Eqs.~\ref{eqt:RingDown22f} and~\ref{eqt:RingDown22q} respectively, allowing the extraction of the ``normalized" spin and mass.

\begin{figure}[!h]
\begin{center}
\includegraphics[trim = 0cm 2cm 18.5cm 0cm, clip, scale=0.48]{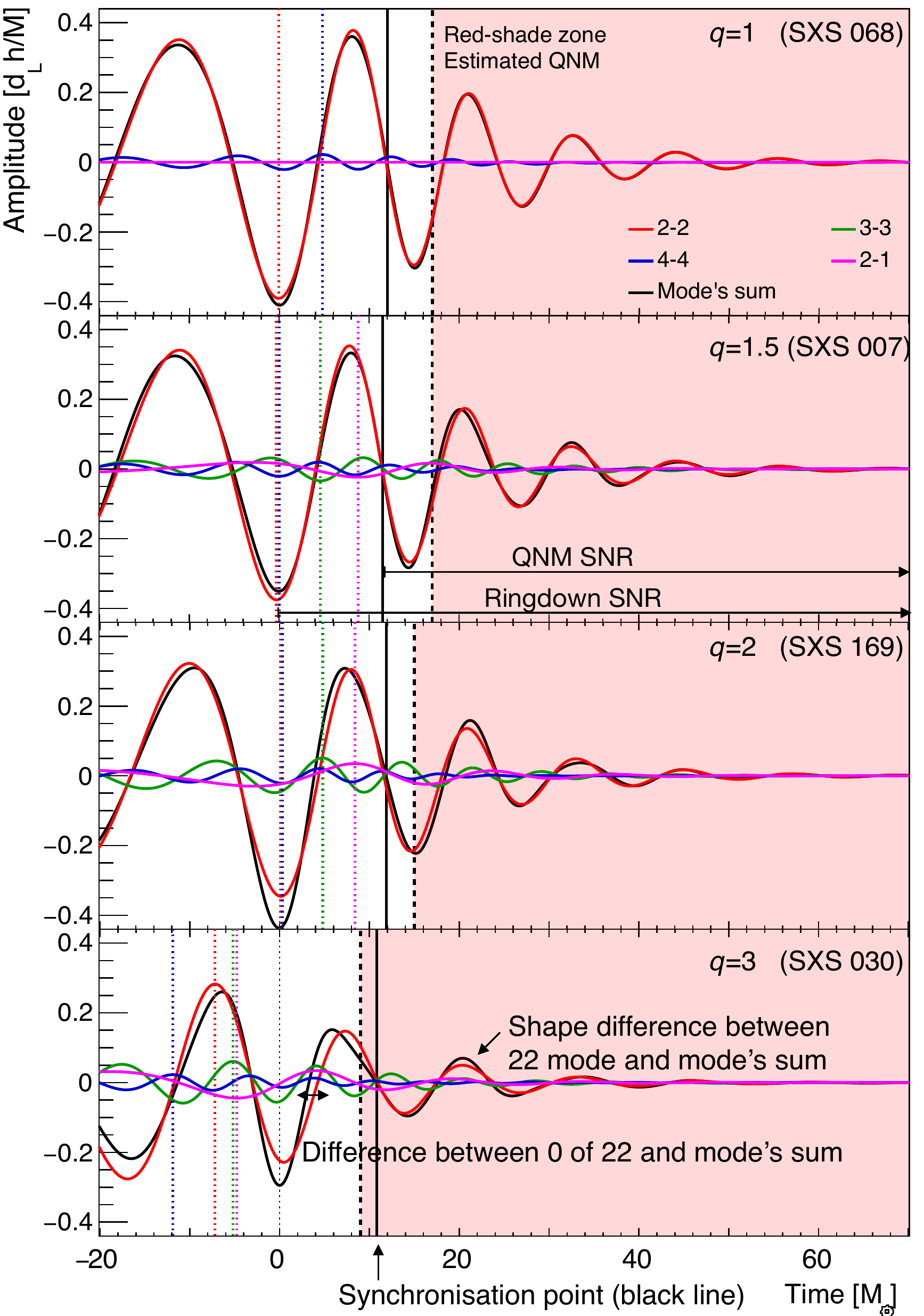}
\end{center} 
\caption{BBH SXS waveforms chosen for our tests: $q\leq3$, non-spinning initial BHs with face-on orientation. The QNMs are: red= (2,2), green=(3,3), blue=(4,4), pink=(2,1) with black representing the QNM sum. We set $t=0$ at the signal maximum amplitude. Short dashed lines mark the maximum of their corresponding mode. The red-shaded zone mark the estimated region of QNMs.  The vertical solid black lines indicate the point from which we synchronize the ringdown signals.} 
\label{fig:ModesOfq3}
\end{figure}

The maximum cumulated SNR is achieved when all signals are perfectly synchronized, which is given by: 
\begin{equation}
max(\mbox{SNR}_{cumulated})=\left(\sum_{i=1}^N \mbox{SNR}_{i}^2 \right)^{1/2}\,,
\label{eq:SNRdef}
\end{equation}
where SNR$_{i}$ is the SNR of each signal and N is the total number of signals. 
We define two SNR's; the QNM SNR, SNR$_{QNM}$, is  defined from the ringdown's synchronisation points; and the ringdown SNR, SNR$_{RD}$, is defined from the peak amplitude (see Fig.~\ref{fig:ModesOfq3}, plot ``$q=1.5$"). SNR$_{RD}$ will be used to select the ringdown signals while SNR$_{QNM}$ is the SNR being accumulated.
Finally, the method efficiency is measured using the ``summation efficiency", SUM$_{eff}$, which corresponds to the ratio between the measured SNR$_{QNM}$ and the maximum expected SNR$_{QNM}$. 

\subsection{Rescaling}
 The rescaling is achieved by resampling the signals according to the ratio $f_{22}(q)/f_{22}(q=1)$. 
 For the current study,
 we derived a fitting function for spinless BHs, by using the SXS metadata, this gives the $f_{22}$ ratio: 
\begin{equation}
\frac{f_{22}(q)}{f_{22}(q=1)}=0.0032 q^2-0.0583q+1.0604 \, .
\end{equation} 
For future studies we will consider spinning initial BHs and adopt the three dimensional function derived in \cite{bib:fitSpinMass}.

As a proof of principle, we use the exact mass ratio.
The error propagation on the mass ratio inferred from the inspiral-merger will be incorporated in future results.
As shown in Tab.~\ref{tab:22MeasuredPeriods}, the periods of the four ringdowns are consistent with one another, showing that the rescaling procedure does not introduce large errors by itself.

\begin{table}[ht]
\caption{Average waveforms periods $<T>$ after rescaling in the QNM linear regime.}
\begin{center}
\begin{tabular}{| l l | ll |}
\hline
Mass ratio &  Average period& &  \\ 
$q$ & $\left<T\right>$ \, [M$_{\odot}$]& &  \\  \cline{1-2} 
1 &  	11.51  &\multicolumn{2}{|c|}{$\left<  \left<T\right> \right>_q=11.55$}  \\
1.5&  11.37  &\multicolumn{2}{|c|}{$ \sigma\left( \left<T\right>\right)_q= 0.20$}\\
2&  	 11.53  &&\\
3&  	 11.84  &&\\ 
\hline
\end{tabular}
\end{center}
\label{tab:22MeasuredPeriods}
\end{table}

\subsection{Synchronization} 
\label{sec:Synchronization}
As indicated in  \cite{bib:Ioannis1}, after the peak GW luminosity, effects of the merging phase are still present in the ringdown. The authors identified the beginning of the QNMs
with the stabilization in time of  the remnant BH frequencies.
 We proceed with similar tests to estimate the QNM starting time. We fit the ringdown waveforms (without noise) at different times using the (2,2) function, Eq.~\ref{eqt:hwave_simple}, and we define the QNM starting time when the spin $a$ becomes constant. The starting times are shown in Fig.~\ref{fig:ModesOfq3} and are compatible with those in \cite{bib:Ioannis1}.

As shown in Fig.~\ref{fig:ModesOfq3}, the QNMs start approximatively one period after the maximum amplitude. 
Though any time after one period can be chosen for synchronization,  ``later" times are disadvantageous due to the quick dampening of QNMs; it is difficult to identify a synchronization point after only one oscillation, while the lower SNR requires more events to extract information.

In order to synchronize the signals, waveform maxima and zeros are easier to identify within noise. As a compromise between a low SNR and influence by the non-linear merger effects, we choose the second zero after the peak amplitude as the synchronization point (see Fig.~\ref{fig:ModesOfq3}). The error due to the merger effect are compared at 3 different times and shown in Tab.~\ref{tab:startingRingdownTimes}.

\begin{table}[ht]
\caption{Relative error on the mass, $M_{BH}$, and spin, $a$, with respect to SXS metadata at 3 different times: the maximum amplitude (t=0), the synchronization point and the estimated beginning of the QNM.  The signal is fitted with the (2,2) mode function, Eq.~\ref{eqt:hwave_simple}.}
\begin{center}
\begin{tabular}{| l | cc | cc |  cc |}
\cline{2-7}
 \multicolumn{1}{c|}{}  &    \multicolumn{6}{|c|}{ Relative errors [\%]}    \\ \hline
Mass &    \multicolumn{2}{|c|}{ $t=0$}    & \multicolumn{2}{|l |}{Synchronisation }    & \multicolumn{2}{l |}{ QNM regime}\\ 
 ratio&    \multicolumn{2}{|c|}{} & \multicolumn{2}{l|}{point  }&\multicolumn{2}{c|}{} \\ 
$q$ &  $a$ & $M_{BH}$&  $a$ & $M_{BH}$& $a$& $M_{BH}$ \\ \hline 
1     &  42  & 42       & 30   & 9  & 15  & 6\\
1.5  &  44  & 40    &  14  & 7 & 12 & 5\\
2     &  47  & 36       & 13   & 6  & 15&5\\ 
3     &  30  & 12      & 6     & 1  & 10& 3\\ \hline 
\end{tabular}
\end{center}
\label{tab:startingRingdownTimes}
\end{table}
 In the QNM regime, we observe systematic errors: $\sim10\%$ higher for the spin $a$ and $\sim5\%$ for the mass, $M_{BH}$, with respect to the  SXS metadata.
Part of the errors in Tab.~\ref{tab:startingRingdownTimes} are also due to the difference between the (2,2) mode, which serves as a fitting function, and the actual GW signal, which is the sum of all the modes (see Fig.~\ref{fig:ModesOfq3}, $q=3$). These errors are expected to be reduced by the summation.\\ 

\begin{figure}[h]
\begin{center}
\includegraphics[trim=0cm 0.0cm 0cm 0cm, clip, scale=0.45]{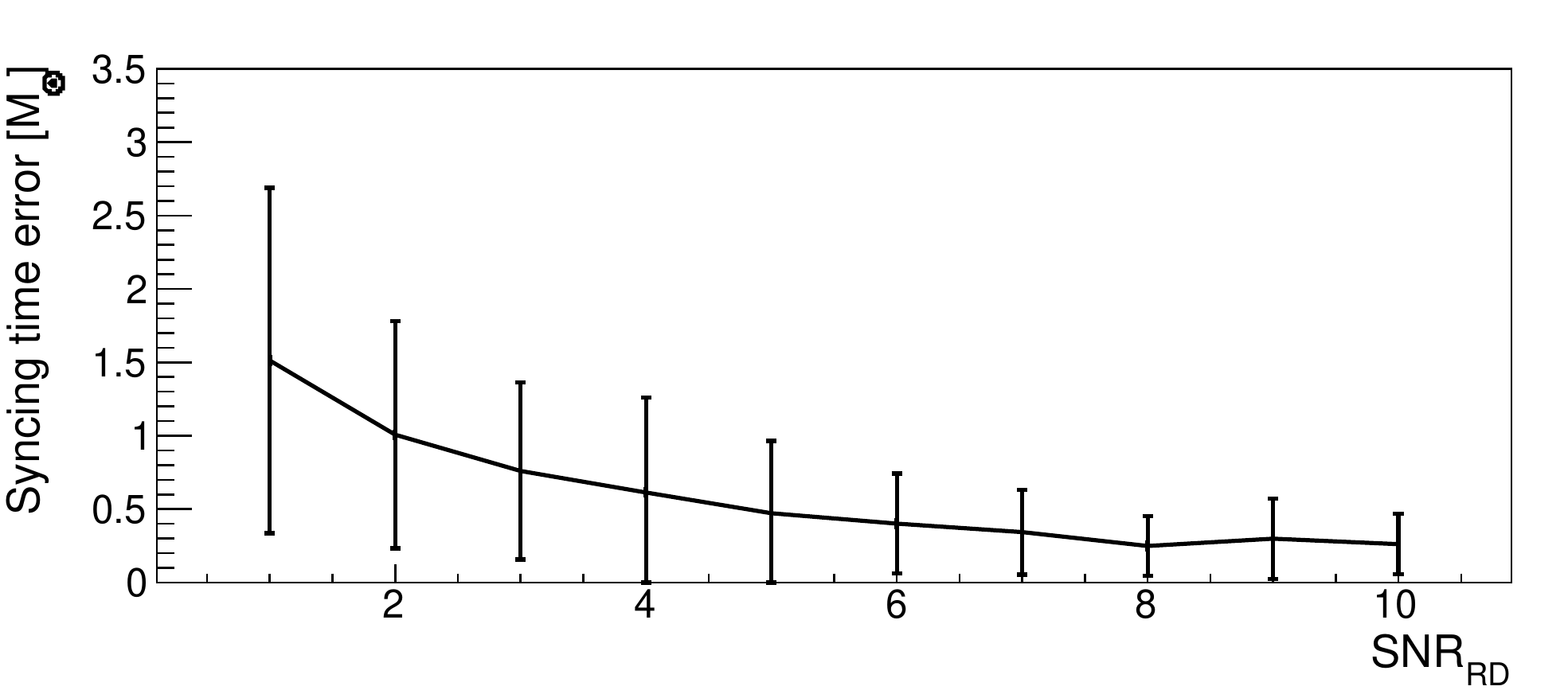}
\end{center}
\caption{Average time shift errors between the (2,2) mode zero and the fitted waveform zero: For each mass ratio and SNR level, the average is computed with 50 injections. Error bars represent standard deviations.}
\label{fig:22-Signal2nd0-noise}
\end{figure}

The synchronization point (2nd zero of waveform after the maximum) is determined by fitting a sine-exponential function, covering a half period, around its expected time,  which, in turn, is 
 estimated by using our knowledge on the maximum amplitude time and expected frequency of the signal.
 The fit is improved by setting its initial parameters, the frequency and damping coefficients, to the values estimated for the rescaling process. The signal is further improved by band-passing it with a narrow window around the mode frequencies. 
The error shift between the (2,2) mode zero and the fitted zero are shown in Fig.~\ref{fig:22-Signal2nd0-noise}. When the SNR$_{RD}\leq1$, the synchronization errors are constrained by the 
implemented limits of the fit. 

\subsection{Subdominant mode perturbations}
The measure of at least two QNMs is required to test the Kerr nature of the remnant \cite{Gossan1}. But presently there is no solution to synchronize the subdominant modes, leaving us with only the dominant (2,2) mode summation.
The subdominant modes are rescaled simultaneously with the (2,2) mode due to the constant ratios between mode frequencies $f_{22}/f_{lm}$.  However, they are not synchronized like the (2,2) modes as
 the phase differences between the (2,2) modes and other modes is different for each $q$. In these conditions, the summed subdominant modes cannot be modeled and thus the divergence they introduce to the (2,2) mode fit cannot be modeled either. The subdominant modes are therefore not considered as extra information but as perturbative noise.

\begin{figure}[h!]
\begin{center}
\includegraphics[angle=-0, scale=0.45]{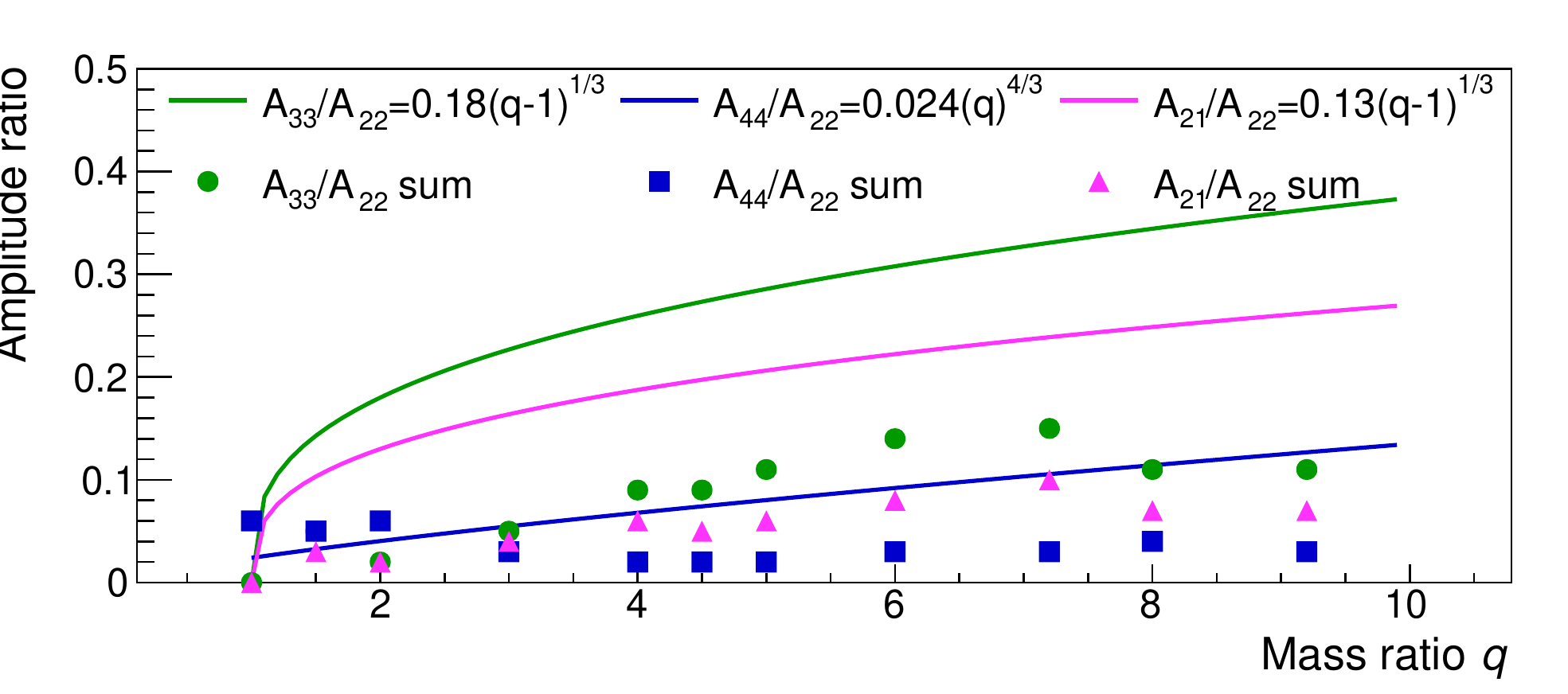}
\end{center} 
\caption{Amplitude ratios between the (2,2) mode and subdominant orders $A_{33}/A_{22}$, $A_{21}/A_{22}$ and $A_{44}/A_{22}$ for different mass ratio $q$. Continuous lines represent amplitude ratio functions previously derived (t=10 M$_\odot$ after the (2,2) peak) \cite{bib:Ioannis1} for single signals. The markers represent the maximum mode ratios when summing all signals from $q=1$ till the indicated $q$. For instance at $q=3$, $q$=1,1.5, 2 and 3 are summed.  } 
\label{fig:AmplitudeRatios}
\end{figure}

Presently, under the BBH face-on condition, the subdominant modes' effect can be neglected when applying the (2,2) mode fit to retrieve the normalized mass, $M^{\prime}_{BH}$ and spin, $a^{\prime}$. 
For $q=3$, the self-imposed upper limit, the ratio with the largest contribution from the subdominant mode is $A_{33}/A_{22}=0.23$ \cite{bib:Ioannis1}; but as the subdominant modes are not summed constructively, the amplitude ratios of the summed signals are lower than for a single signal, see Fig.~\ref{fig:AmplitudeRatios}. 

The BBH inclination changes the relative amplitude between the modes as shown by Eqs 20-23 in \cite{bib:Ioannis1}. 
The sub-dominant mode (2,1) can reach up to $\sim$0.6 times the dominant mode (2,2) for the case of q =3 and edge-on system (this is the highest mode contribution). The contribution of (2,1) mode is still low as compared to the (2,2) mode for the mass ratios considered. For such particular case the systematics will introduce higher error as (2,2) and (2,1) are at the same frequency. Other effects such as mode-mixing could also arise leading to higher errors  \cite{bib:decodingQNM}. In future work, we will consider the change due the BBH inclination.

In the case of single signals, the subdominant mode affect the synchronization point, changing its time. Their effect is proportional to the subdominant modes' amplitudes, thus increasing with $q$ as shown in Fig.~\ref{fig:ModesOfq3}. The largest synchronization shift, $t_s$, is achieved when they are in phase with each other but not with the (2,2) mode maximum. In this scenario, the synchronization points shift between the (2,2) mode zero and the ringdown zero by $t_s=\{ 0.4, 0.8, 0.9,1.5\}$ M$_{\odot}$ for $q=\{1, 1.5, 2,3\}$. The error introduced is comparable to the time shift due to noise at SNR $=4$ (see Fig.~\ref{fig:22-Signal2nd0-noise}).\\

\subsection{Cumulated SNR and parameter extraction}
\begin{figure}[ht!]
\begin{center}
\includegraphics[ trim=0cm 0.7cm 0cm 0.1cm, clip, scale=0.45]{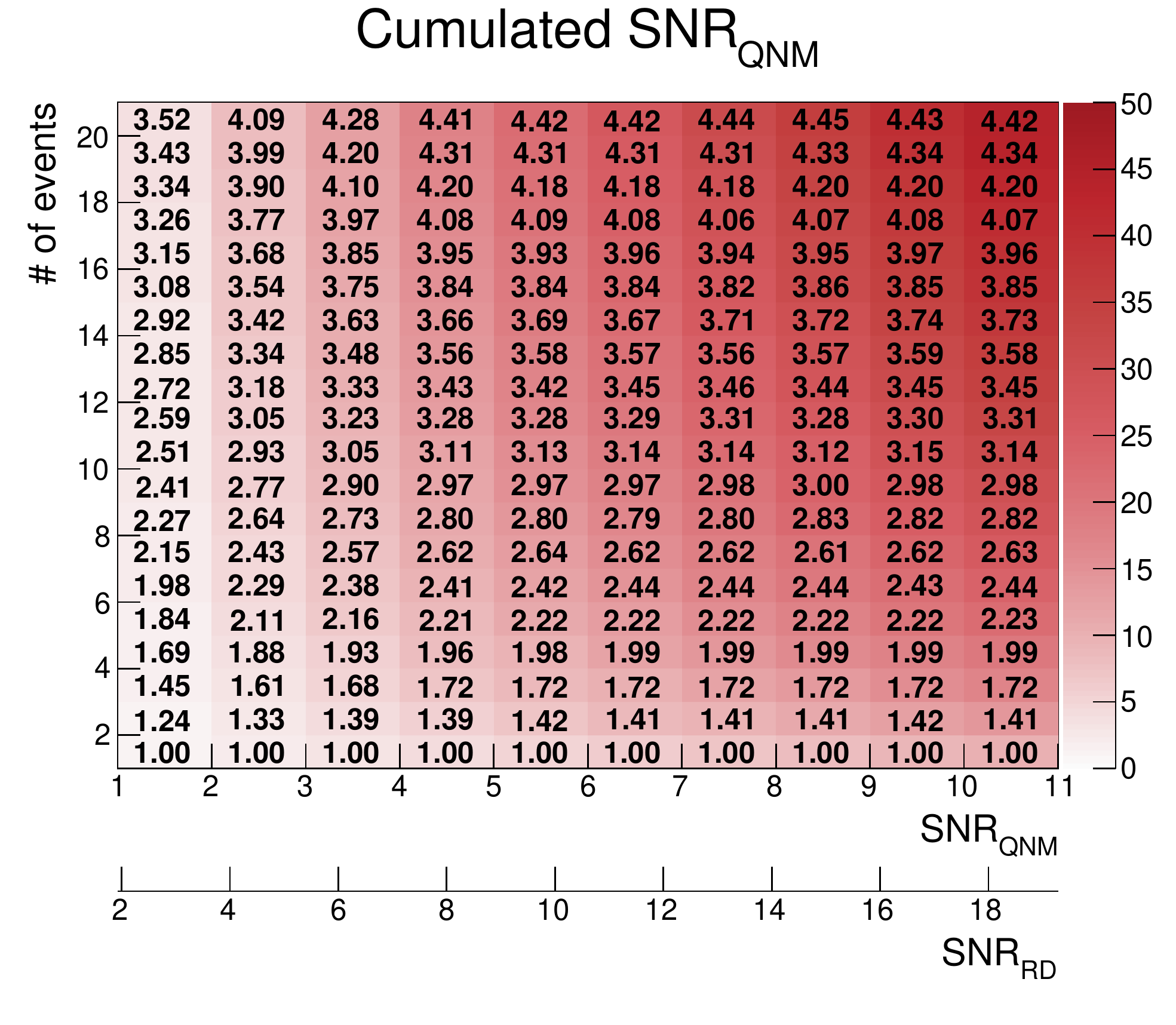}
\caption{Cumulated SNR$_{QNM}$ from $N$ event signals with the same original SNR$_{QNM}$:  For each entry of the table, the test is repeated 50 times with waveforms chosen randomly between the 4 mass ratios. The color scale indicates the average cumulated SNR$_{QNM}$ and the written numbers correspond to the ratio between injected and the averaged cumulated SNR$_{QNM}$. \\
The  SNR$_{RD}$ scale is shown for indication about the SNR being used for synchronization. }
\label{fig:FitSNR1}
\end{center}
\end{figure}

We proceed to inject our four signals into white noise, each with 10 different SNR$_{QNM}$.  In Fig.~\ref{fig:FitSNR1}, for each SNR$_{QNM}$, we sum up, incrementally, 20 randomly sampled signals with the previous synchronization method (repeated waveforms are used). 
The cumulated SNR, according to Eq.~\ref{eq:SNRdef}, should increase by a factor $\sqrt{N}$. For low SNR$_{RD}$,  the synchronization errors are higher (see Fig.~\ref{fig:22-Signal2nd0-noise}), therefore the signals are not summed constructively, and the ratios shown in the low SNR$_{QNM}$ columns of Fig.~\ref{fig:FitSNR1} are lower than $\sqrt{N}$.

\begin{figure}[ht!]
\begin{center}
\includegraphics[trim=0cm 2.8cm 0cm 0.2cm, clip, scale=0.45]{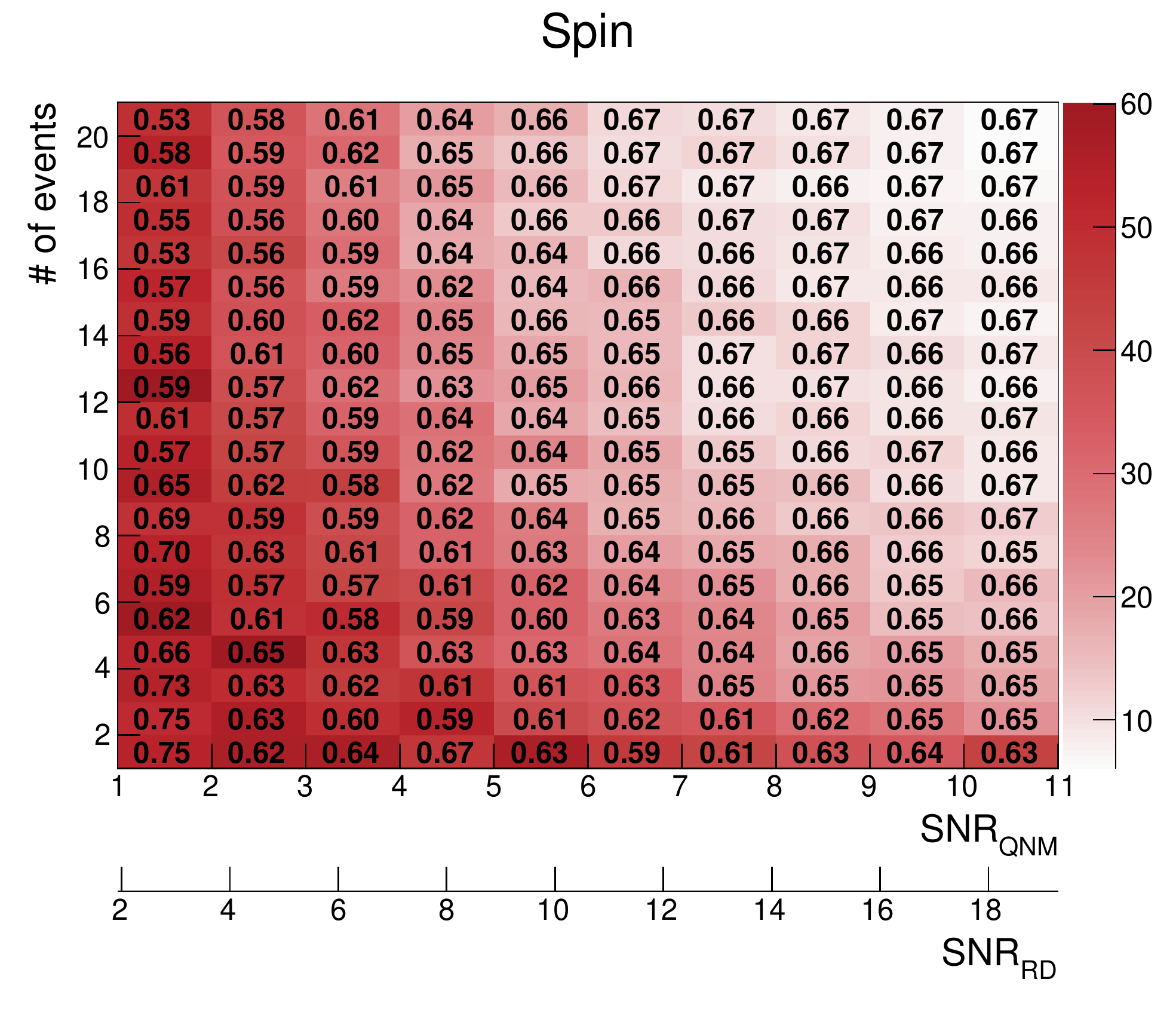} 
\includegraphics[trim=0cm 2.8cm 0cm 0.3cm, clip, scale=0.45]{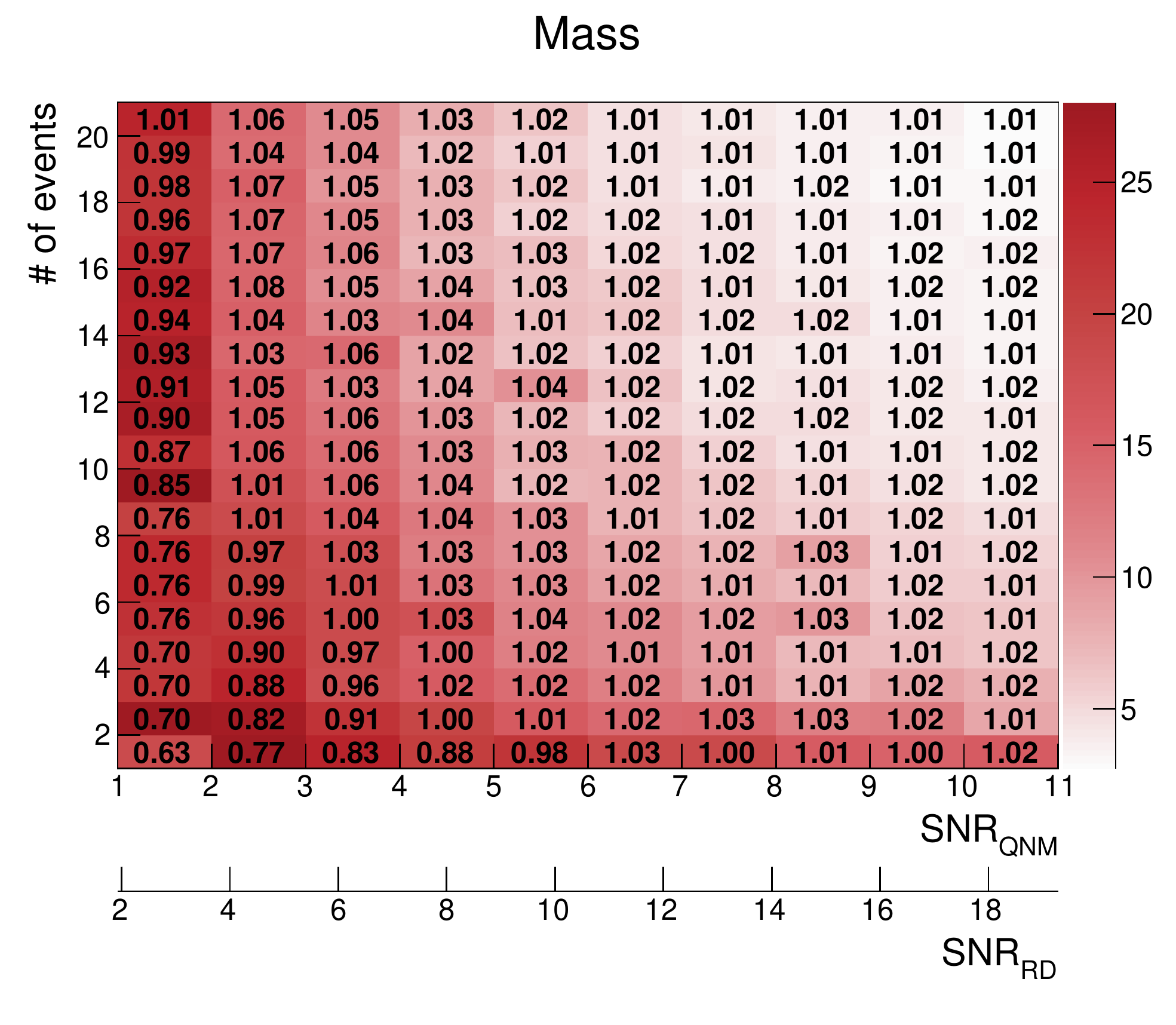}
\caption{ Average and standard deviations of the fitted spin and masse values from $N$ event signals with the same original SNR$_{QNM}$:: The color scale indicates the standard deviation in \% while the written numbers correspond to the average (each test is repeated 50 times). The waveforms are chosen randomly between the 4 mass ratios and the resultant signal is fitted with the (2,2) mode functions Eqs.~\ref{eqt:RingDown22f}-\ref{eqt:RingDown22q}.}
\label{fig:Fits}
\end{center}
\end{figure} 
The precision on the normalized mass and spin as a function of the SNR$_{QNM}$ and the number of summed signals are shown in Fig.~\ref{fig:Fits}. 
The expected spin and mass values of the resulting signal from the four randomly sampled waveforms are inferred by fitting the signal sum without noise. The fit results are $a^{\prime}= 0.66$ and $M_{BH}^{\prime}$=1.0~[$M_{BH}$/$M_{BBH}$] (The SXS remnant mass are given proportionally to the BBH initial total mass). These values are affected by the aforementioned errors in the previous section (e.g. propagation of non-linear mergers effects in the ringdown, signal synchronization, discrepancy between the (2,2) mode and the actual GW signal) which explains why $M_{BH}\nless1$. 

With a collection of low SNR$_{QNM}$ signals, the standard deviation is up to 60\% on the spin and is 30\% on the mass. 
This, however, obviously improves for the higher SNR$_{QNM}$, e.g. 10 signals with SNR$_{QNM}=3$ gives 35\% and 15\% precision on the spin and mass respectively.
Except for small variations, the precision follows this cumulated SNR$_{QNM}$ trend. For values of signals with SNR$ge$10, the precision is compatible with the ones predicted in \cite{bib:Echev_fits}.

\section{Application to a population of simulated ringdowns}
\label{sec:application}
\begin{figure}[ht!]
\begin{center}
\includegraphics[trim=0cm 0cm 0cm 0.5cm, clip, scale=0.45]{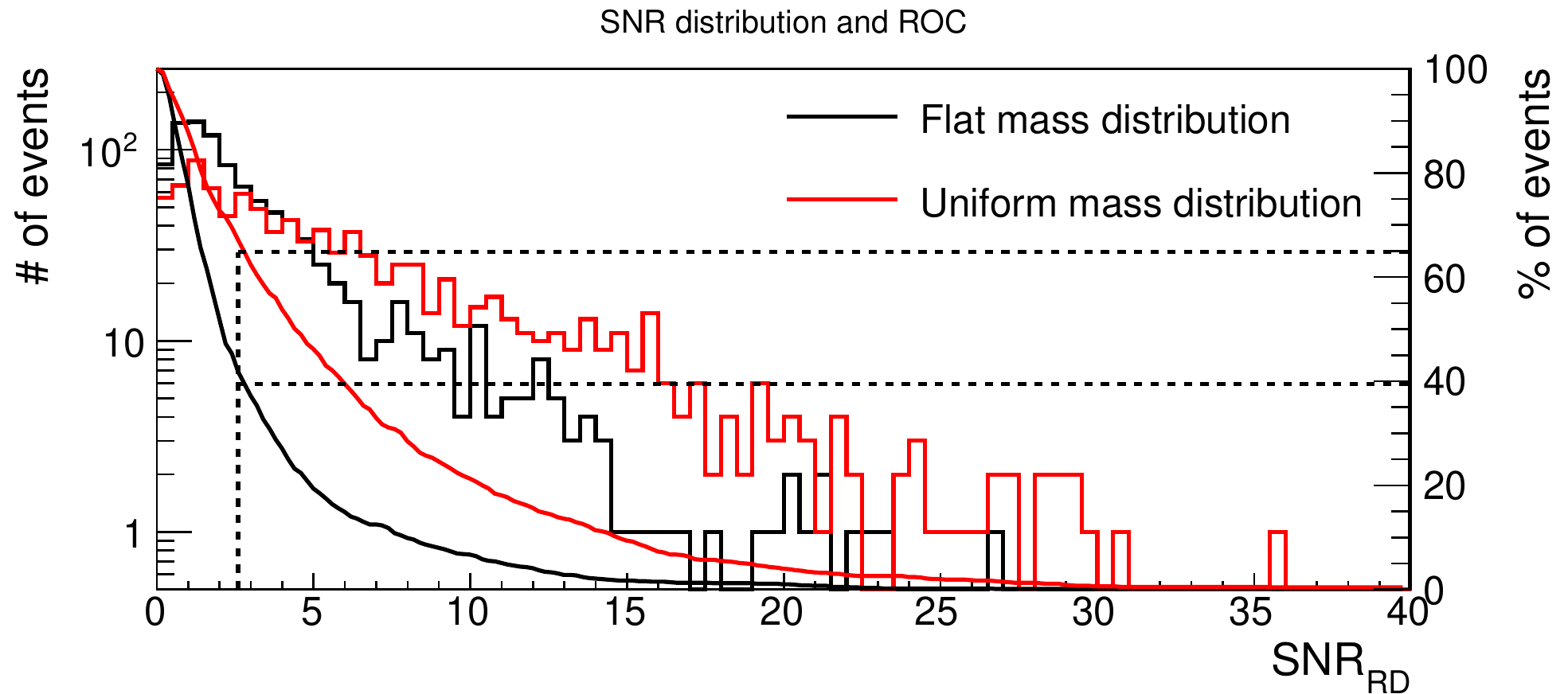}
\caption{The histograms represent the number of events with the respective SNR$_{RD}$ for each BBH mass distribution (black for flat and red for uniform).  The receiver operating characteristic (ROC) for the two mass distributions is represented by the solid lines (number of events against SNR$_{RD}$) while the SNR$_{RD}$ threshold is highlighted with the dashed line.\\
}
\label{fig:allSNRdist}
\end{center}
\end{figure}

The main interest of the summation method is to retrieve physical information from the weak SNR ringdown, SNR$_{RD}$, signals. 
In order to understand how many events (of our restricted BBH type) could be employed in our analysis, using SEOBNR \cite{bib:SEOBNR}, we simulate the merger signals of two BBH populations.  Each has 1000 events, with different mass distributions:  \textit{uniform} distribution in component masses and \textit{flat} in $\log (m1 )$ and $\log (m2 )$. 
The BBHs are uniformly distributed in volume and with a total mass between 10-100~M$_{\odot}$. In Fig.~\ref{fig:allSNRdist} their SNR$_{RD}$ values are given for the designed sensitivity of the advanced interferometers LIGO and VIRGO \cite{bib:TheSensitiveOfAdvance}. In the following tests, for each population the 4 waveforms are randomly sampled and injected into noise with a random value SNR$_{RD}$ from the distributions shown in Fig.~\ref{fig:allSNRdist}.

\begin{figure}[ht!]
\begin{center}
\includegraphics[trim=0cm 0cm 0cm 0cm, clip, scale=0.45]{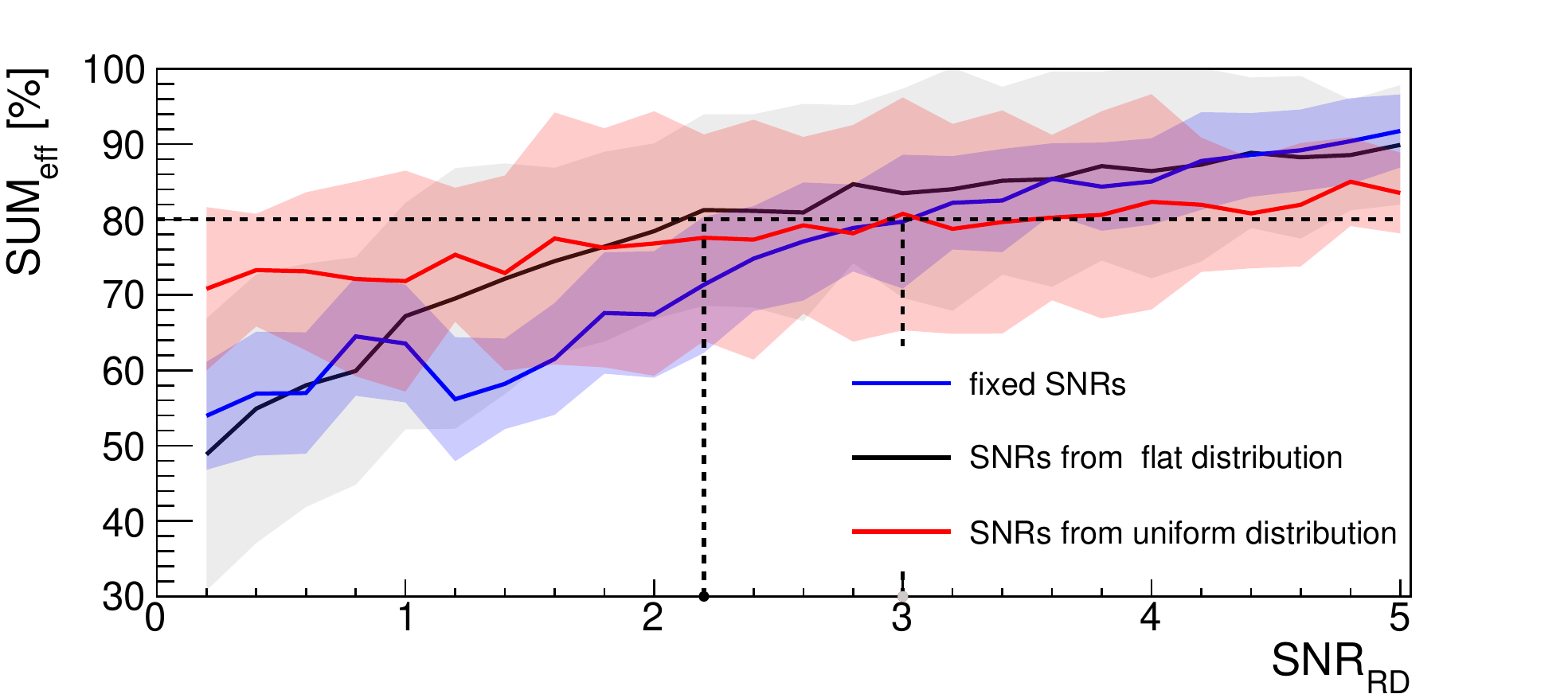}
\caption{SNR summation efficiency, SUM$_{eff}$, as a function of different SNR thresholds for 3 SNR$_{RD}$ distributions. The blue line corresponds to simulations with SNRs equal to the threshold, i.e. fixed.  The red and black lines are the SNRs of the uniform and flat populations respectively. For each entry,  20 event signals are summed and the test is repeated 20 times; the color bands represent the standard deviations.}
\label{fig:SelectionSNRcuts}
\end{center}
\end{figure}

As we have seen in Section \ref{sec:Synchronization}, the synchronization errors, which prevent the signals being summed constructively, are worse for lower SNR$_{RD}$.  It would therefore be beneficial to introduce a SNR$_{RD}$ threshold, which will allow us to select those events with usable SNR$_{RD}$.
In Fig.~\ref{fig:SelectionSNRcuts}, the SUM$_{eff}$ as a function of different SNR$_{RD}$ thresholds is shown. The results for 3 SNR$_{RD}$ distributions (the two mass distribution and a limit case with SNR$_{RD}$ equal to the threshold) are compared with 80\% efficiency; this value was chosen as the curves stabilize above it.  The flat mass distribution has the highest SNR$_{RD}$ from its detected signals and its curve passed the 80\% efficiency at a lower threshold, SNR$_{RD}= 2.2$, than the other distributions; their curves reach the 80\% efficiency at SNR$_{RD}=3$. These efficiency values lie between a non-constructive summation of 20 signals, 47\% ($N^{-1/4}$), and fully constructive, 100\%. 
The 50\% efficiency at very low SNR$_{RD}$ ($\sim0.2$) is a method artifact; the fit covers a region where the signal's zero is expected, thus all synchronization times are close to the real ones.
 
\begin{figure}[ht!]
\begin{center}
\includegraphics[trim=0cm 1.2cm 0cm 0cm, clip, scale=0.45]{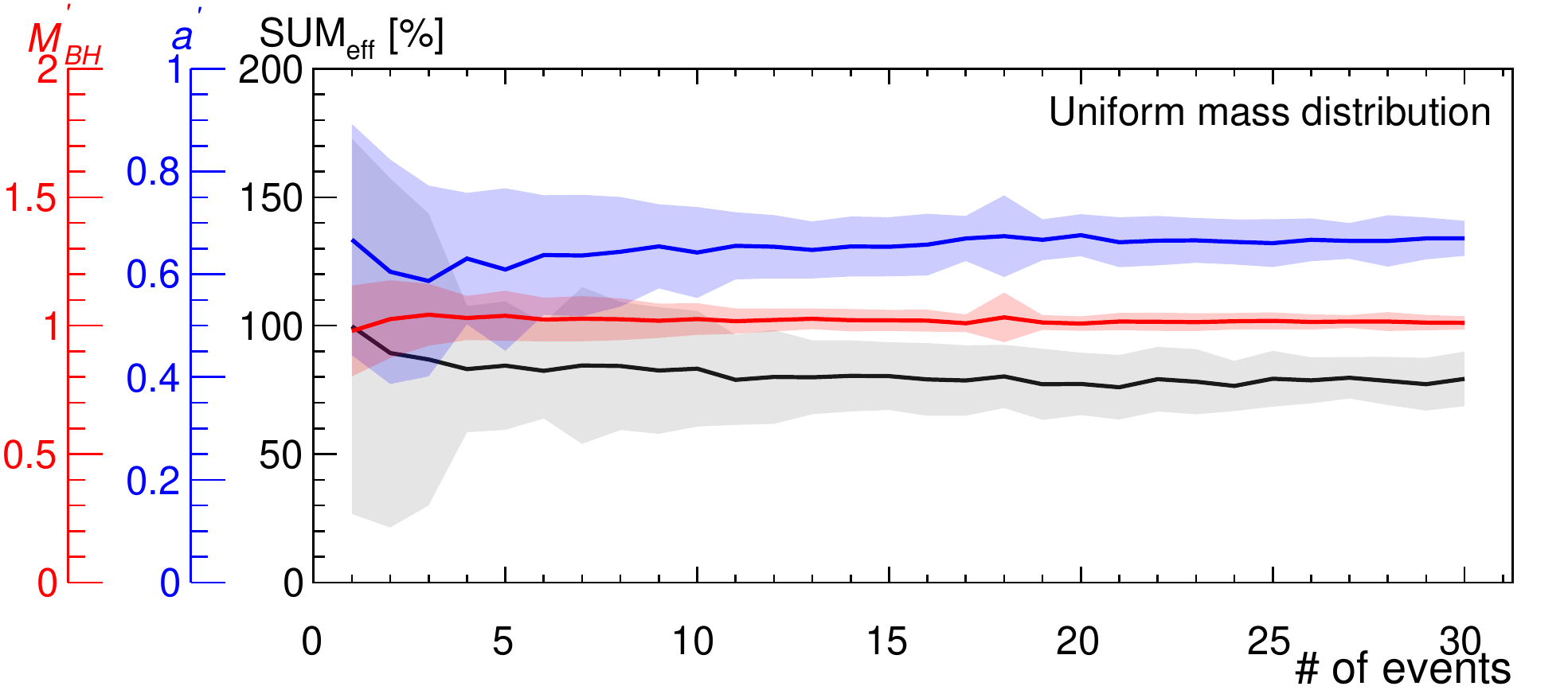} 
\includegraphics[trim=0cm 0cm 0cm 0.7cm, clip, scale=0.45]{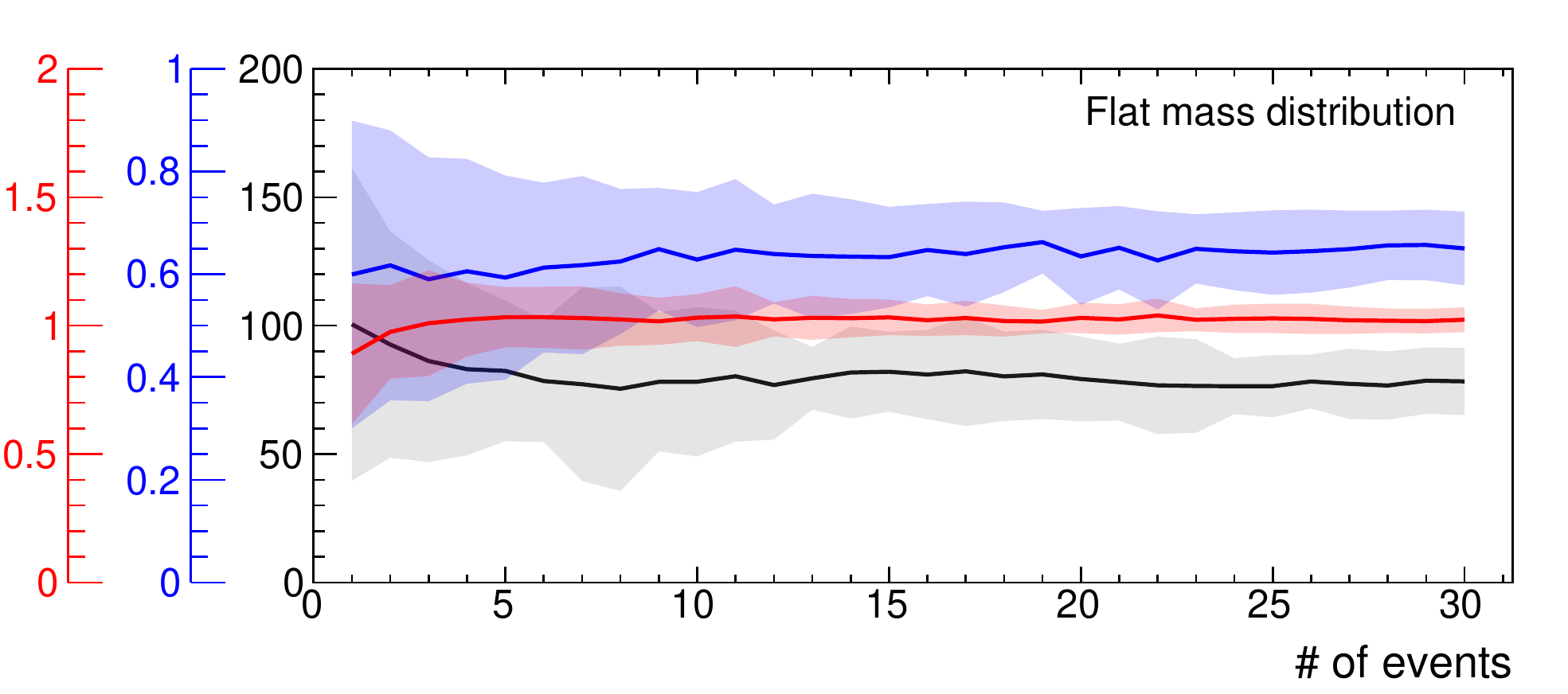} 
\caption{ Average values of mass, spin and SUM$_{eff}$ for different numbers of event signals summed. The expected values are: $M_{BH}^{\prime}=1$, $a^{\prime}=0.66$, and SUM$_{eff}=80$\%. The average is computed with 20 repetitions, and the color bands represent the standard deviations.}
\label{fig:SNRSpinMass}
\end{center}
\end{figure}
Taking into account standard deviations of the 3 curves, we choose a threshold  SNR$_{RD}\ge2.6$. Depending on the expected BBH mass distribution, 40\% to 70\% of the signals will be selected (see Fig.~\ref{fig:allSNRdist}).

In Fig.~\ref{fig:SelectionSNRcuts}, the SUM$_{eff}$ is computed for 20 summed signals. In Fig.~\ref{fig:SNRSpinMass}, the SUM$_{eff}$, the mass and the spin are shown for several numbers of summed signals after applying the chosen threshold, SNR$_{RD}=2.6$. 
The average of the SUM$_{eff}$ stabilizes around 80\% due to our threshold choice and its effect on 
synchronization.  The standard deviation for a few events is large because it depends directly on the SNR$_{RD}$ distributions; while for more events, this effect is averaged. In addition, the standard deviations of the flat mass distribution are still larger due to its bulkier distribution at low SNR$_{RD}$.  The precision on the mass and spin in both cases follow the SNR trend.


\section{Conclusion}
\label{sec:Conclusion}
In this paper, we tested the signal summation technic that we developed for the (2,2) mode. Between many technical difficulties, we focus, for this first study, on the signal synchronisation. We show that by selecting signals with SNR$_{RD}\ge2.6$,  we can ensure a signal summation efficiency of 80\%. For the selected BBH types, depending on the expected BBH mass distribution,  40 to 70\% of the potential BBH signals detected can be used to extract the remnant properties.

The extracted information gives statistical information on the remnant BH population. The use and limitations of this information still need to be studied; but before this step, the method should undergo improvements by expanding it to a wider BH parameter space. As a first study, we overviewed and identified briefly many limitations that need to be addressed.  Future work will involve initial spinning BHs as well as different BH orientations (not face-on). 

We are also working on new and more robust signal synchronization techniques to allow the use of more signals.  Finally, an important step will be the synchronization of the subdominant modes. Their information will allow a more definitive test on the Kerr nature of the remnant BH population. 

\section*{Acknowledgments}
This work has been supported by NSF grant PHY 1505308. S. Tiwari is supported by the People Program (Marie Curie Actions) of the European Union FP7/2007-2013/ (PEOPLE-2013-ITN) under REA grant agreement No. [606176].\\
We would like to acknowledge with much appreciation the discussions with A.~Heffernan, B. F. Whiting  and L. London. We acknowledge the comments and references that C. Lousto gave us.  We also thank the SXS collaboration, M. Boyle and H. Pfeiffer, for kindly answering our questions. 

\appendix
\section*{Appendix}

\subsection{Quasi Normal Modes Gravitational Waves}
\label{sec:QNM}
The QNM Gravitational Waves are given by: 
\begin{equation}
 h=\sum_{nlm} {}_{-2}Y_{lm}(\iota, \phi) h_{nlm} \,,
\label{eqt:hwave}
\end{equation}
where $_{-2}Y_{lm}$ are the spin-weighted spherical harmonics. The index $n$ is the overtone while $l$ and $m$, are the spheroidal harmonic indices. The angles are given in the source frame: $\iota$ is the angle between the system spin and the line of sight, and $\phi$ is the azimuth angle. The GW amplitudes of the  QNMs are defined by:

\begin{multline}
 h_{nlm} = A_{nlm} \frac{M_{BH}}{r} e^{-\pi f_{nlm}/Q_{nlm} t}\\ \times \sin(2\pi f_{nlm} t +\Phi_{nlm})\,,
 \label{eqt:hwave2}
\end{multline}
where $A_{nlm}$ and $\Phi_{nlm}$ are respectively the amplitude and phase of each mode. They depend on physical phonemes happening inside the BH and will be provided by simulations. $M_{BH}$ is the remnant BH's mass, and $r$ is the distance to the source.
Frequencies $f_{nlm}$ and quality factors $Q_{nlm}$ are given by \cite{bib:Leaver85}, 
\begin{eqnarray}
f_{nlm}=\frac{1}{2\pi} \frac{c^3}{G M_{BH}}\left[f_1+f_2 (1-a)^{f_3}\right] \label{eqt:Fnlm} \,\\
Q_{nlm}=q_1+q_2 (1-a)^{q_3}\, .\label{eqt:Qnlm}\,
\end{eqnarray}
where $a=[0,1]$ is the dimensionless spin; $a=0$ implies non-spinning while $a=1$ corresponds to the maximum spin, where the innermost stable circular orbit (ISCO) is close to the BH radius.

Several authors \cite{bib:PressTeukolsky73, bib:Detweiler, bib:Schutz_Will, bib:Leaver85}  have developed semi-analytic or numerically methods to computes the QNMs. Nowadays, the parameters $f_1, f_2, f_3, q_1, q_2$ and $q_3$ are determined by fitting simulation results \cite{Berti2006, Berti2009}, the modes $(l,m)=$(2,2), (3,3), (2,1) and (4,4) are then given by, 
 \begin{eqnarray}
f_{22} = \frac{1}{2\pi} \frac{c^3}{G M_{BH}}\left[1.525 - 1.157 (1-a) ^{0.129}\right] \label{eqt:RingDown22fp}\\ 
Q_{22} = 0.700+1.419 (1-a)^{-0.499} \label{eqt:RingDown22p}
\end{eqnarray}
\begin{eqnarray}
f_{21} = \frac{1}{2\pi} \frac{c^3}{G M_{BH}}\left[0.600- 0.234 (1-a) ^{0.418}\right] \label{eqt:RingDown21f}\\ 
Q_{21} = -0.300+2.356 (1-a)^{-0.228} \label{eqt:RingDown21}
\end{eqnarray}
\begin{eqnarray}
f_{33} = \frac{1}{2\pi} \frac{c^3}{G M_{BH}}\left[1.896 - 1.304 (1-a) ^{0.182}\right] \label{eqt:RingDown33f}\\ 
Q_{33} = 0.900+2.343 (1-a)^{-0.481} \label{eqt:RingDown33}
\end{eqnarray}
\begin{eqnarray}
f_{44} = \frac{1}{2\pi} \frac{c^3}{G M_{BH}}\left[2.300 - 1.505 (1-a) ^{0.224}\right] \label{eqt:RingDown44f}\\ 
Q_{44} = 0.700+1.419 (1-a)^{-0.483} \label{eqt:RingDown44}
\end{eqnarray}

\subsection{Tested waveforms}
\label{sec:waveforms}
 All the SXS waveforms used for our tests are listed in Tab.~\ref{tab:waveforms}; they are low eccentricity and non-spinning initial BHs. 
For the ringdown, we use a specific set of data called ``outermost", where NR extractions were performed without extrapolation. This set better describes the ringdown as extrapolations will contain numerical errors.

\begin{table}[ht]
\caption{Chosen simulations from SXS. 
All values are expressed in geometrical units,$c=G=1$, the time is normalized by the total initial mass $M_{BBH}$ while the initial masses are normalized to $M_{BBH}=1$.}
\begin{center}
\begin{tabular}{|l |  c c c|}
\hline
Mass ratio $q$ &  Waveform Id. & Mass $M_{BH}$ & Spin $a$\\  \hline 
1 & 002  &  0.952 & 0.622 \\ 
1.5 &  007& 0.955 & 0.606 \\ 
2 & 169 &  0.961 & 0.576 \\ 
3 & 030 &  0.971 & 0.510 \\ 
4 & 167&  0.978 & 0.451 \\ 
4.499 & 190  & 0.980 & 0.425\\ 
5 & 054 &  0.982 & 0.402  \\ 
6 & 166 &  0.985 & 0.362\\ 
7.187& 188  & 0.988 & 0.323  \\ 
8 & 063 & 0.989 & 0.300 \\ 
9.167& 189  & 0.990 & 0.273  \\ \hline
\end{tabular}
\end{center}
\label{tab:waveforms}
\end{table}


\bibliography{./MyBIB.bib}

\end{document}